# JointNET: A Deep Model for Predicting Active Sacroiliitis from Sacroiliac Joint Radiography

Sevcan Turk[1,2]*, Ahmet Demirkaya[3,4,5], M Yigit Turali[4,5], Cenk Hepdurgun[2], Salman UH Dar[4,5], Ahmet K Karabulut[2], Aynur Azizova[6], Mehmet Orman[9], Ipek Tamsel[2], Ustun Aydingoz[6], Mehmet Argin[2,7], Tolga Çukur[4,5,8]

1. Department of Radiology, University of Michigan, MI/USA
2. Department of Radiology, Ege University, Izmir/Turkey
3. Department of Electrical and Computer Engineering, Northeastern University, MA/USA
4. National Magnetic Resonance Research Center (UMRAM), Bilkent University, Ankara/Turkey
5. Department of Electrical and Electronics Engineering, Bilkent University, Ankara/Turkey
6. Department of Radiology, Hacettepe University, Ankara/Turkey
7. EMOT Hospital, Izmir, Turkey
8. Neuroscience Program, Aysel Sabuncu Brain Research Center, Bilkent University, Ankara/Turkey
9. Department of Biostatistics, Ege University, Izmir/Turkey

*Correspondence to:

Sevcan Turk, MD
Department of Radiology
University of Michigan
1500 E. Medical Center Dr, B1G503A
Ann Arbor, MI 48109-5030
sevcant@med.umich.edu
sevcanturk.ege@outlook.com

Manuscript Type: Original Research
Funding: None

**Summary Statement:** We developed JointNet to evaluate active inflammation on sacroiliac joint radiographs using MRI as gold standard with a comparison of human observers. JointNet's mean accuracy was 90.2% (95% CI: 87.6%, 92.8%) which outperformed experienced radiologists (accuracy 54-64%).

**Key Points:**

- Deep learning algorithms can detect active inflammation on sacroiliac joint radiographs with a 90.2% mean accuracy when using MRI as gold standard.

- JointNet's success to predict active inflammation outperformed experienced radiologists. The sensitivity was 69.0% (95% CI :65.3%, 72.7%) and specificity 90.4% (95% CI :87.8 % 92.9%) whereas the Radiologists' sensitivities were 46% and 65%, and the specificities were 57% and 64%.

- Radiographs are used as first line imaging modality to evaluate SIJ. Further image analysis using convolutional neural networks may provide additional information and may help to reduce MRI need for the evaluation of SIJ inflammation.

**Abstract**

Purpose: To develop a deep learning model that predicts active inflammation from sacroiliac joint radiographs and to compare the success with radiologists.

Materials and Methods: A total of 1,537 (augmented 1752) grade 0 SIJs of 768 patients were retrospectively analyzed. Gold-standard MRI exams showed active inflammation in 330 joints according to ASAS criteria. A convolutional neural network model (JointNET) was developed to detect MRI-based active inflammation labels solely based on radiographs. Two radiologists blindly evaluated the radiographs for comparison. Python, PyTorch, and SPSS were used for analyses. P<0.05 was considered statistically significant.

Results: JointNET differentiated active inflammation from radiographs with a mean AUROC of 89.2 (95% CI:86.8%, 91.7%). The sensitivity was 69.0% (95% CI :65.3%, 72.7%) and specificity 90.4% (95% CI :87.8 % 92.9%). The mean accuracy was 90.2% (95% CI: 87.6%, 92.8%). Positive predictive value was 74.6% (95% CI: 72.5%, 76.7%) and negative predictive value was 87.9% (95% CI: 85.4%, 90.5%) when prevalence was considered 1%. Statistical analyses showed a significant difference between active inflammation and healthy groups (p<0.05). Radiologists' accuracies were less than 65% to discriminate active inflammation from sacroiliac joint radiographs.

Conclusion: JointNET successfully predicts active inflammation from sacroiliac joint radiographs, with superior performance to human observers.

**Introduction**

It has been more than a decade since MRI entered the center stage of the Assessment of SpondyloArthritis international Society (ASAS) guidelines, which stipulates that active sacroiliitis on MRI or definite sacroiliitis on radiographs according to modified New York criteria is a major criterion for the diagnosis of axial spondyloarthritis [1]. Inflammatory lesions are visible on MRI before structural changes are visually detectable on radiography. Although MRI can also show structural damage lesions such as erosions, subchondral sclerosis, fat deposition in the periarticular bone marrow, and ankylosis, periarticular osteitis or bone marrow edema is a prerequisite on MRI for the diagnosis of active sacroiliitis. However, interobserver reliability is variable, with moderate agreement at best for the presence of active sacroiliitis and erosions [2].

Conventional radiography remains the first-line imaging investigation in patients suspected of having inflammatory arthritis of all kinds. Radiographs, however, primarily reflect bony damage and capture a projectional view of the imaging volume. Visual inspection of such images by human observers often lacks the sensitivity to detect early disease or incremental changes in the short term. In assessments of structural lesions consistent with chronic sacroiliitis on MRI, radiographs show a sensitivity ranging from 25% to 75%, and a specificity ranging from 71% to 91% across observers [3]. Furthermore, the misclassification rate ranges from 9% to 23% for negative radiographs with positive MRIs. Notably, the interobserver agreement is weak for radiography interpretation by expert radiologists [4]. According to the modified New York criteria, five stages of radiographic changes in the sacroiliac joints (SIJs) can be identified [5].

In this study, we hypothesized that active inflammation during sacroiliitis causes quantitatively detectable changes on radiographs, even in cases where human observers have difficulty picking up relevant signatures. To test this hypothesis, we set out to develop a novel deep learning model based on an ensemble architecture that combines residual and dense convolutional backbones for predicting SIJ inflammation, as confirmed by the current gold standard of MRI.

**Materials and Methods**

**Dataset:** Institutional IRB approval and informed consent were obtained for this retrospective study. Radiographs showing modified New York criteria grade 0 SIJs whereby an MRI was available within 1.5 months of radiography acquisition were reviewed from 2010 to 2020 at PACS archives of two university hospitals. A total of 1,537 (augmented 1752) grade 0 SIJs of 768 patients were retrospectively analyzed. Radiographs were grouped into two categories: 1) Patients with active inflammation on MRI; 2) patients without any findings of sacroiliitis on MRI (Figure 1). A total of 330 joints (augmented 720) had active inflammation on MRI according to ASAS criteria. Patients with orthopedic implants or evidence of operation at or nearby the SIJs were excluded from the study.

**Image Acquisition:** Sacroiliac joint radiographs with Ferguson views were obtained using digital devices (Samsung Ceiling Digital X-ray GC85A, Seoul, Korea and Axiom Aristos, Siemens Healthcare Solutions, Erlangen Germany) at single radiology units within two tertiary referral hospitals. Four units, a 3T or a 1.5T systems (Siemens Verio and Siemens Amira, Erlangen, Germany or Signa, GE Healthcare, Milwaukee, WI, USA; Symphony Tim or Aera, Siemens Healthcare Solutions; Achieva or Ingenia, Philips Medical Systems, Best, The Netherlands) within the same hospitals were used for SIJ MRI acquisitions with oblique-coronal STIR and pre-and post-contrast oblique-coronal T1W and oblique-axial fat-saturated T1W images. In the rare instances when i.v. contrast was not given, fat-saturated T1W sequence on the transverse oblique plane was substituted with fat-saturated T2W or STIR sequence.

**Data Review:** All SIJ MRIs were reviewed by two dedicated musculoskeletal radiologists (one with 23 years and the other with 20 years of experience) blindly, and the cases were classified into "active sacroiliitis" and "normal SIJ" groups. A musculoskeletal radiologist with 5 years of experience and a neuroradiology trained radiologist with 3 years of experience reviewed the radiographs blindly. SPSS was used for cross tables and chi-square analysis.

**Preprocessing:** Radiographs were first split around the midline horizontally to separate left and right joints. This procedure enabled to analyze the inflammation status of each joint individually. As there may be correlated physiology between each pair of joints, joints within a given patient were either assigned to the training or the validation set. To further focus on the critical morphology, an automated region-of-interest (ROI) extraction was employed via template matching with a 224x224 size [6]. To normalize signal intensity variations across images, adaptive histogram equalization was performed [7] [Supporting Figure 1]. Data augmentation was implemented using PyTorch torchvision [8] and OpenCV [9] libraries. Supporting Table 1-2 validation analyses demonstrate the important of data augmentation and normalization on detection performance.

**JointNET.** A deep learning model was developed to detect MRI-based active inflammation labels given sacroiliac (SI) radiographs as well as age and gender information as two separate channels. JointNet leveraged an ensemble architecture with dense and residual convolutional backbones. To alleviate data requirements while improving model performance, transfer learning was employed where the dense and residual backbones were initialized with pre-trained weights from ImageNet [10]. An overview of the JointNet was given in Figure 2. Each backbone processes radiographs augmented with age and gender features prior to the output layer as additional information. The outputs from the dense and residual backbones are aggregated via averaging to produce the final output of JointNet.

**Model Training and Evaluation.** Model training and testing were implemented in Python and PyTorch. Assessments were performed via a 10-fold cross-validation procedure. In each fold, data were randomly split into 92% training (1592 radiographs) and 8% testing (160) sets. The test datasets didn't overlap across the folds. The training was performed by minimizing cross-entropy loss via the AdamW optimizer for 20 epochs and learning rate $5 \times 10^{-3}$. Test performance was assessed by the following quantitative metrics: area under the curve (AUC), accuracy, sensitivity, specificity, positive predictive value (PPV), negative predictive value (NPV), precision, recall, and F1 score.

Five different convolutional backbones were considered to construct the ensemble in JointNet. For this purpose, the popular DenseNet121 [11], ResNet152 [12], AlexNet [13], VGG11 [14], and InceptionV3 [15] backbones were considered. These models were first individually examined to identify promising candidates, and ensemble configurations from pairs of models were then formed.

**Statistical Procedures.** SPSS and Python was used for statistical assessments. $P<0.05$ was considered statistically significant. Descriptive statistics for success rate and Kappa correlation coefficient between observers and Chi-Square were used for analysis. AUCs of various ROC curves were compared with Wilcoxon Sign Rank Test.

**Results**

The mean age was 42 for both genders [51% (391/768) female and 49% (377/768) male]. Radiologists spent approximately 3 seconds per image evaluation. Kappa correlation between observers to diagnose normal radiographs, inflammation on radiographs and overall were

0.43, 0.27 and 0.38 respectively. There were no objective decision-making findings on radiographs. Radiologists' accuracies (Reader 1 and Reader 2) were 54% (830/1537) and 64% (984/1537) with a sensitivity of 46% (202/436) and 65% (284/436), specificity of 57% (628/1101) and 64% (701/1101), NPV of 73% (628/862) and 82% (701/853), and PPV of 30% (202/675) and 41% (284/684) (Supporting table 6-7-8).

Five backbone architectures were individually trained to discriminate active inflammation on radiographs (DenseNet, ResNet, AlexNet, VGG, Inception). Supporting Table 1, 2 and 3 lists demonstrate the importance of data augmentation, data normalization and auxiliary age and gender information to detection performance. The top-performing backbone was DenseNet ($p<0.05$; Wilcoxon's signed-rank test)(Supporting Table 4). It achieved a mean AUC of 88.9 (95% CI:87.1%, 90.8%). The sensitivity was 70.9% (95% CI :67.7%, 74.1%), the specificity was 87.5% (95% CI :84.3%, 90.7%). The mean accuracy was 87.3% (95% CI: 84.1%, 90.5%). Positive predictive value was 75.1% (95% CI: 73.2%, 77.1%) and negative predictive value was 85.3% (95% CI: 82.1%, 88.4%) when prevalence was considered 1% (Supporting Figure 2).

Next, candidate pairs of convolutional blocks were evaluated to implement the ensemble architecture in JointNet. The combination of the dense and residual blocks achieved the highest performance among all methods ($p<0.05$), so it was used to implement JointNet. Accordingly, JointNet successfully differentiated active inflammation using radiographs ($p<0.05$). It achieved a mean AUC of 89.2 (95% CI:86.8%, 91.7%). The sensitivity was 69.0% (95% CI :65.3%, 72.7%) with a specificity of 90.4% (95% CI :87.8 % 92.9%). The mean accuracy was 90.2% (95% CI: 87.6%, 92.8%). Positive predictive value was 74.6% (95% CI: 72.5%, 76.7%) and negative predictive value was 87.9% (95% CI: 85.4%, 90.5%) when prevalence was considered 1% (Table 1)(Supporting Figure 3,4). The corresponding histogram output of class probabilities for healthy and active inflammation cases are shown in Figure 3. Only 8 out of 80 total cases with active inflammation were missed, and 24 out of 80 healthy cases were misclassified. These results suggest that JointNet is highly sensitive for active inflammation cases.

**Discussion**

This study showed that invisible, subtle histogram changes on SIJ radiographs suggesting active inflammation could be evaluated with deep learning algorithms. The proposed JointNet model based on an ensemble of residual and dense convolutional blocks could distinguish active inflammation on SIJ radiographs with a success level superior to radiologists' observations. We also showed that adding age and gender as clinical input increased the AUC and sensitivity which correlates with prior clinical studies [20].

Bressem et al. built a deep learning algorithm to discriminate sacroiliitis from regular patients using SIJ radiographs with AUC 94 [16]. However, they included all grades of the disease and did not analyse active inflammation. Shenkman et al. used lumbar CT scans for automatic grading of sacroiliitis, whereby the Combined Random Forest and CNN-based algorithm's mean AUC was 97 to differentiate sacroiliitis from the control group [17].

Chen et al. showed that bone marrow edema detection on CT was possible for sacroiliitis by using dual-energy CT [18]. They evaluated 40 patients' virtual non-calcium images with a comparison of sacroiliac MRI. Using a -44 HU cutoff value resulted in 76.9% sensitivity and 91.5% specificity when compared to radiologists' observations [18].

In another study Bressem et al. classified inflammatory and structural changes on MRI with 94 and 89 AUC using deep learning tools [19]. However, they did not use radiographs or grading.

We included only New York criteria grade 0 SIJs to the evaluation which limits the generalizability of the results to all grades. And no external validation set was available. Further research is obviously needed with higher-grade sacroiliitis.

**Conclusion**

This study showed that subtle density changes surrounding the SIJ with active inflammation not detectable by the human eye but picked up by automated deep learning algorithms to successfully discriminate sacroiliitis from healthy controls. This proof of concept may eventually pave the way for radiographs to replace costly MRI studies.

**Acknowledgement**

**Figures/Tables:**

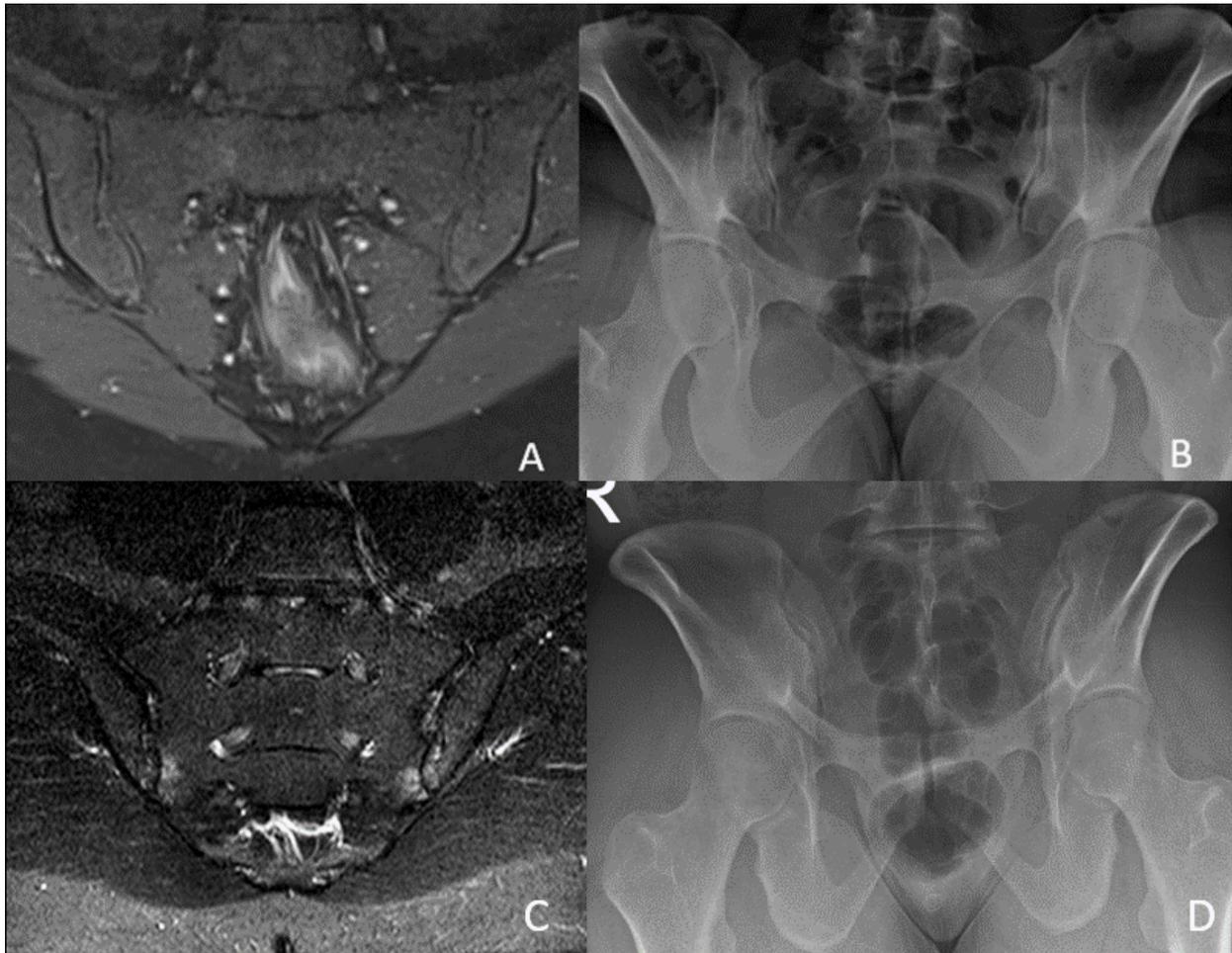

Figure 1: Postcontrast fat-saturated T1-weighted image of SIJ without active inflammation (A). SIJ radiograph of the same patient, classified as modified New York criteria grade 0 (B). Another patient with left SIJ focal contrast enhancement on postcontrast fat-saturated T1-weighted image consisted of inflammation (C). No significant visual abnormality on SIJ radiograph, classified as modified New York criteria grade 0 (D).

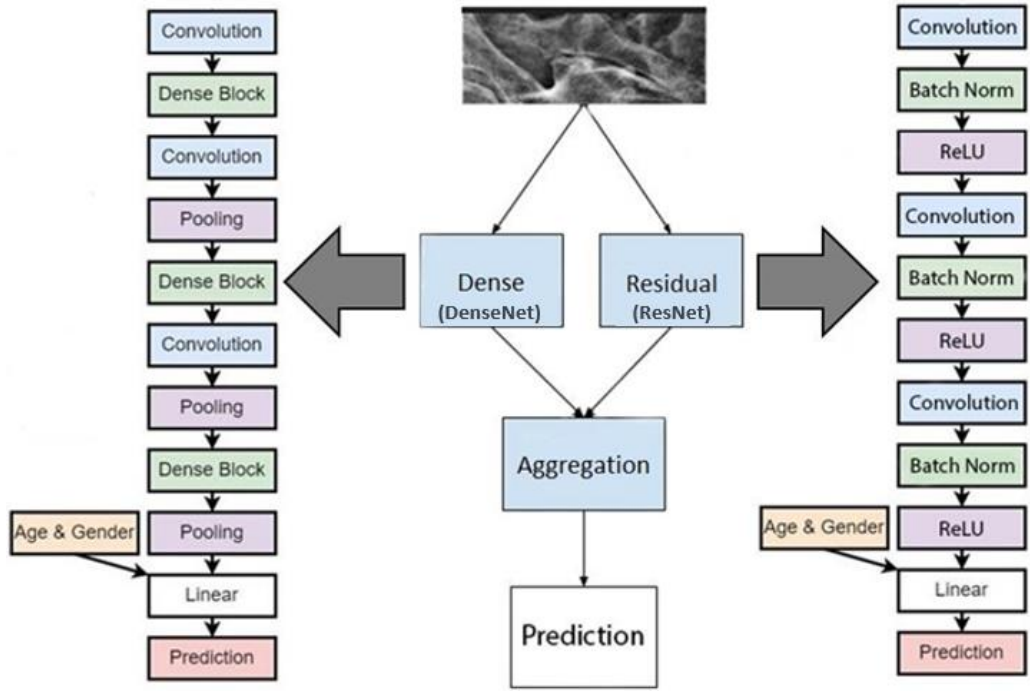

Figure 2: A deep learning model was developed to detect MRI-based active inflammation labels given sacroiliac (SI) radiographs as well as age and gender information as two separate channels. JointNet leverages an ensemble architecture with dense and residual convolutional backbones. Each backbone processes radiographs augmented with age and gender features prior to the output layer. The outputs are aggregated via averaging to produce the final output of JointNet.

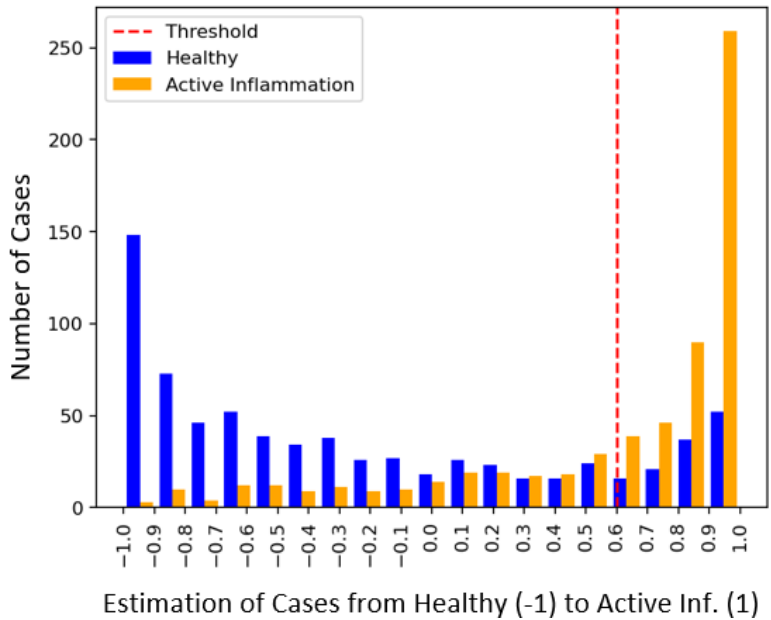

Figure 3: A histogram of test radiographs separately shown for healthy (blue) and active inflammation (orange) cases according to confirmed diagnoses. The horizontal axis indicates the detection probability of active inflammation based on the output of JointNet.

| Metrics | Value | 95% Confidence Interval |
|---|---|---|
| AUC | 0.8920 | 0.868 - 0.917 |
| Sensitivity | 0.6900 | 0.653 - 0.727 |
| Specificity | 0.9040 | 0.878 - 0.929 |
| Positive Predictive Value | 0.7460 | 0.725 - 0.767 |
| Negative Predictive Value | 0.8790 | 0.854 - 0.905 |

Table 1: Area Under the Curve (AUC), Accuracy, Sensitivity, Specificity, Positive Predictive Value (PPV), And Negative Predictive Value (NPV) metrics for JointNet.

**Supporting Material:**

|  | AUC |
|---|---|
| w. Augmentation | 0.8252 ± 0.0497 |
| w/o. Augmentation | 0.7548 ± 0.0501 |

Supporting Table 1: Five convolutional backbones were individually trained, with and without data augmentation on original radiographs. For these analyses, data normalization was not considered. Mean ± STD of AUC is shown for the average performance across the individual backbones.

|  | AUC |
|---|---|
| w. Normalization | 0.8242 ± 0.0389 |
| w/o. Normalization | 0.7687 ± 0.0342 |

Supporting Table 2: Five convolutional backbones were individually trained, with and without data normalization during pre-processing of radiographs. For these analyses, data augmentation was not considered. Mean ± STD of AUC is shown for the average performance across the individual backbones.

| Model | AUC | Sensitivity | Spesificity | PPV | NPV |
|---|---|---|---|---|---|
| JointNet | 0.8921 | 69.00% | 90.37% | 74.58% | 87.94% |
| JointNet w/o Gender | 0.8746 | 63.00% | 90.13% | 70.96% | 86.69% |
| JointNet w/o Age | 0.8735 | 64.13% | 89.38% | 71.73% | 85.95% |

Supporting Table 3: JointNet was compared against ablated variants without age and without gender information on subjects. Mean AUC, sensitivity, specificity, positive predictive value (PPV) and negative predictive value (NPV) are listed.

| Model | AUC | Sensitivity | Spesificity | PPV | NPV |
|---|---|---|---|---|---|
| DenseNet-ResNet | 0.89210 | 69.00% | 90.38% | 74.58% | 87.94% |
| DenseNet-Inception | 0.89100 | 68.75% | 89.25% | 74.19% | 86.51% |
| DenseNet-VGG | 0.87920 | 65.00% | 89.25% | 71.90% | 85.86% |
| DenseNet-AlexNet | 0.87580 | 62.75% | 89.88% | 70.89% | 86.21% |

Supporting Table 4: Area Under the Curve (AUC), Accuracy, Sensitivity, Specificity, Positive Predictive Value (PPV), and Negative Predictive Value (NPV) metrics for ensemble models.

| Model Comparison | p-value |
|---|---|
| DenseNet - ResNet | $3.6943 \times 10^{-5}$ |
| DenseNet - Alexnet | $2.1231 \times 10^{-11}$ |
| DenseNet - VGG | $1.1617 \times 10^{-4}$ |
| DenseNet - Inception | $1.0000 \times 10^{-2}$ |

Supporting Table 5: The resulting p-values of Wilcoxon Sign Rank Test between different backbone models. The test shows that the null hypothesis is rejected since all p-values is smaller than 0.05, therefore DenseNet is statistically different than the other backbone models.

| READER | Accuracy | Sensitivity | Specificity | NPV | PPV |
|---|---|---|---|---|---|
| Reader 1 | 54% | 46.3% | 57% | 72.9% | 29.9% |
| Reader 2 | 64.1% | 65.1% | 63.7% | 82.2% | 41.5% |

Supporting Table 6: Accuracy, sensitivity, specificity, NPV, and PPV among the readers.

| Crosstab Reader 1 | Negative for inf/Gold | Positive for inf/ Gold | Total |
|---|---|---|---|
| Negative for inf | 628 | 234 | 862 |
| Positive for inf | 473 | 202 | 675 |
| Total | 1101 | 436 | 1537 |

Supporting Table 7: Cross table for Reader 1

| Crosstab Reader 2 | Negative for inf/Gold | Positive for inf/ Gold | Total |
|---|---|---|---|
| Negative for inf | 701 | 152 | 853 |
| Positive for inf | 400 | 284 | 684 |
| Total | 1101 | 436 | 1537 |

Supporting Table 8: Cross table for Reader 2

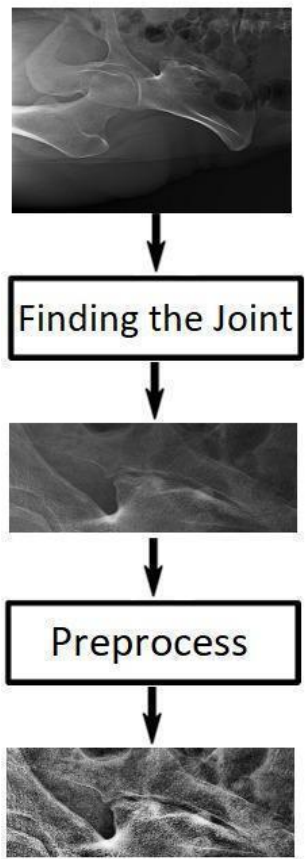

Supporting Figure 1: Radiographs were split around the midline horizontally to separate the left and right joints. To further focus on the critical morphology, an automated extraction module was employed that employ template matching to identify a region-of-interest (ROI). Adaptive histogram equalization was used on the ROI to normalize signal intensity variations across images.

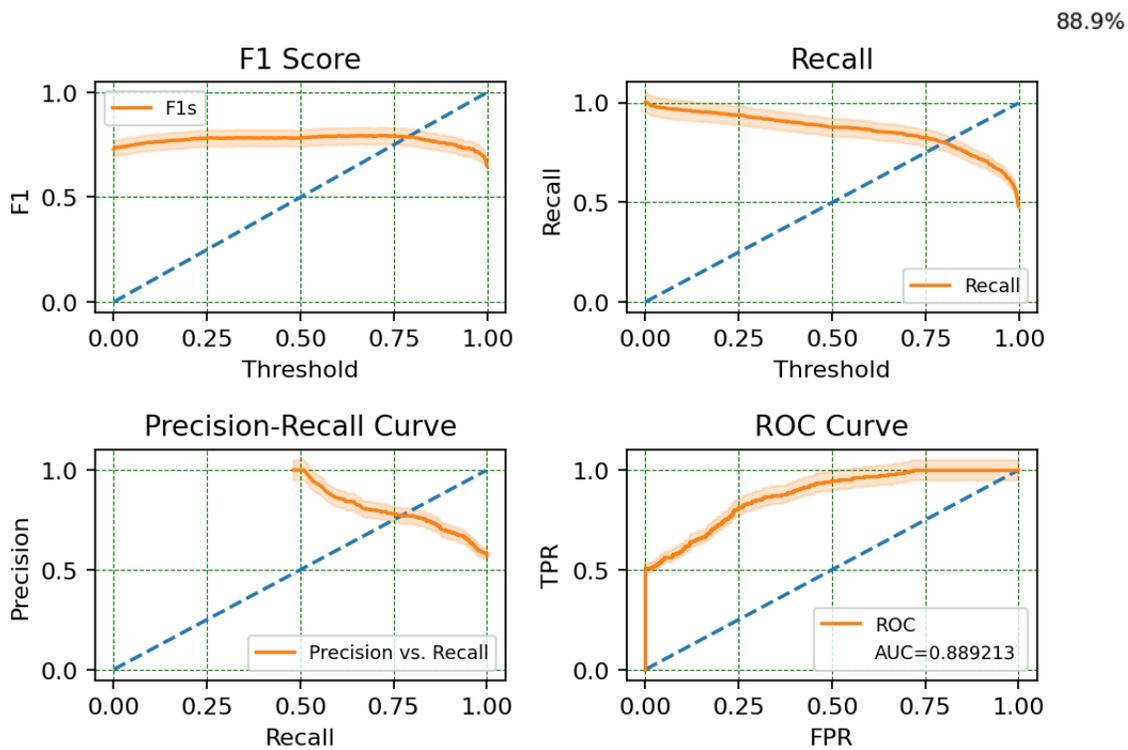

Supporting Figure 2: DenseNet differentiated active inflammation from radiographs with a mean AUROC of 88.9 (95% CI:87.1%, 90.8%). The sensitivity was 70.9% (95% CI :67.7%, 74.1%) with a specificity of 87.5% (95% CI :84.3%, 90.7%). The mean accuracy was 87.3% (95% CI: 84.1%, 90.5%). Positive predictive value was 75.1% (95% CI: 73.2%, 77.1%) and negative predictive value was 85.3% (95% CI: 82.1%, 88.4%) when prevalence was considered 1%.

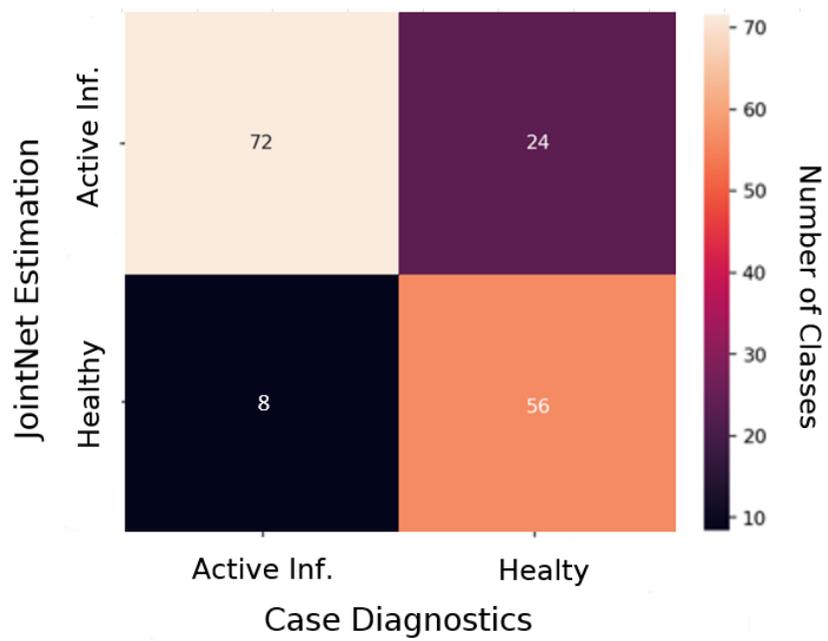

Supporting Figure 3: The confusion matrix for the JointNet model on a collection of 160 test samples. The horizontal axis shows the JointNet predicted class of given samples, whereas the vertical axis shows the actual diagnosis.

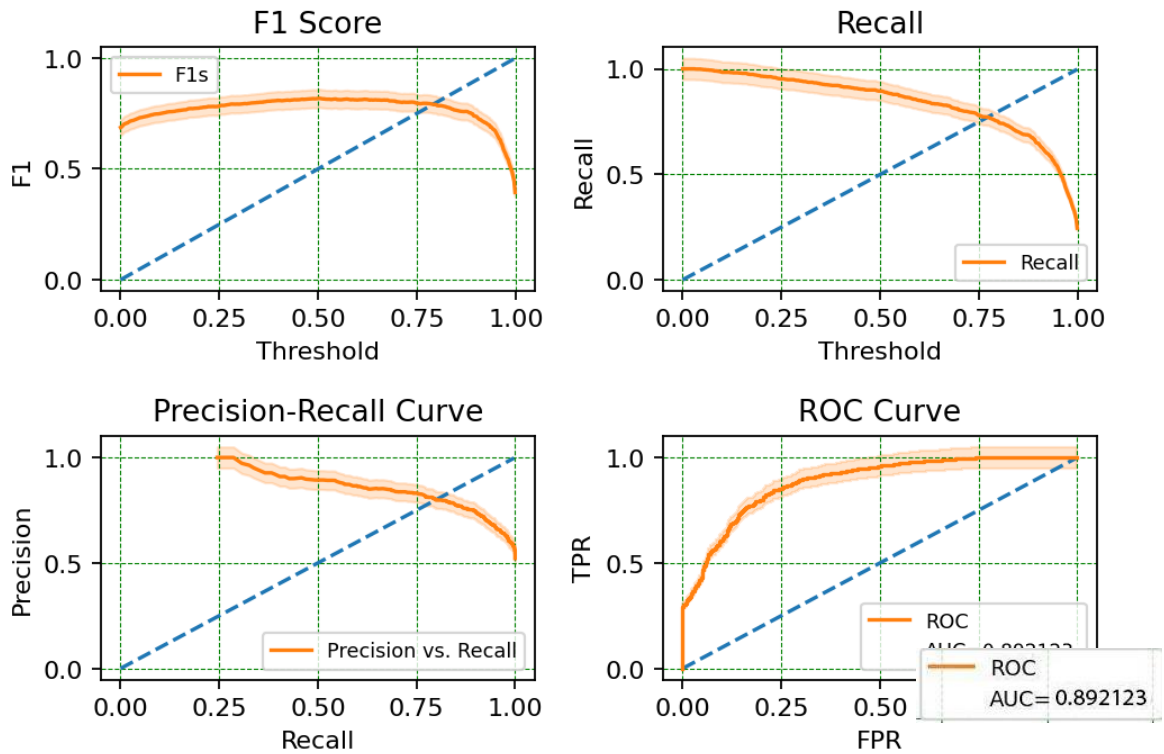

Supporting Figure 4: The F1 score and Recall curves as a function of decision threshold, along with Precision-Recall and ROC curves shown for JointNet. The orange curves denote JointNet performance, whereas the dashed blue lines denote the performance of a random classifier.

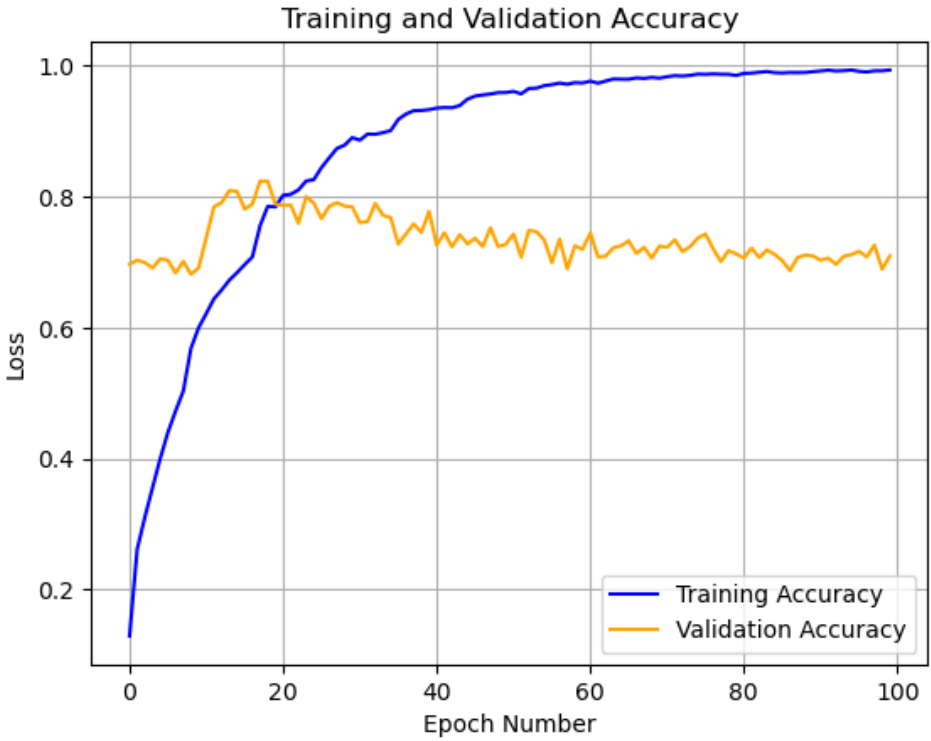

Supporting Figure 5: Train/Validation accuracy curves of JointNet model. Since Overtraining starts around epoch number 20, the model was trained until epoch number 20.

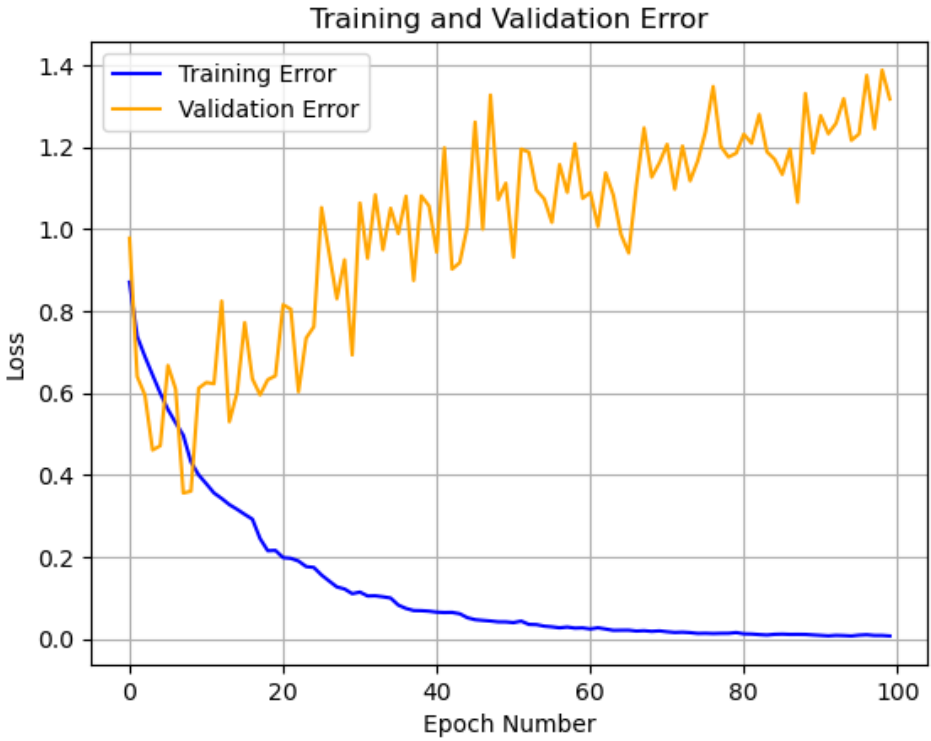

Supporting Figure 6: Train/Validation loss curves of JointNet model. Since Overtraining starts around epoch number 20, the model was trained until epoch number 20.

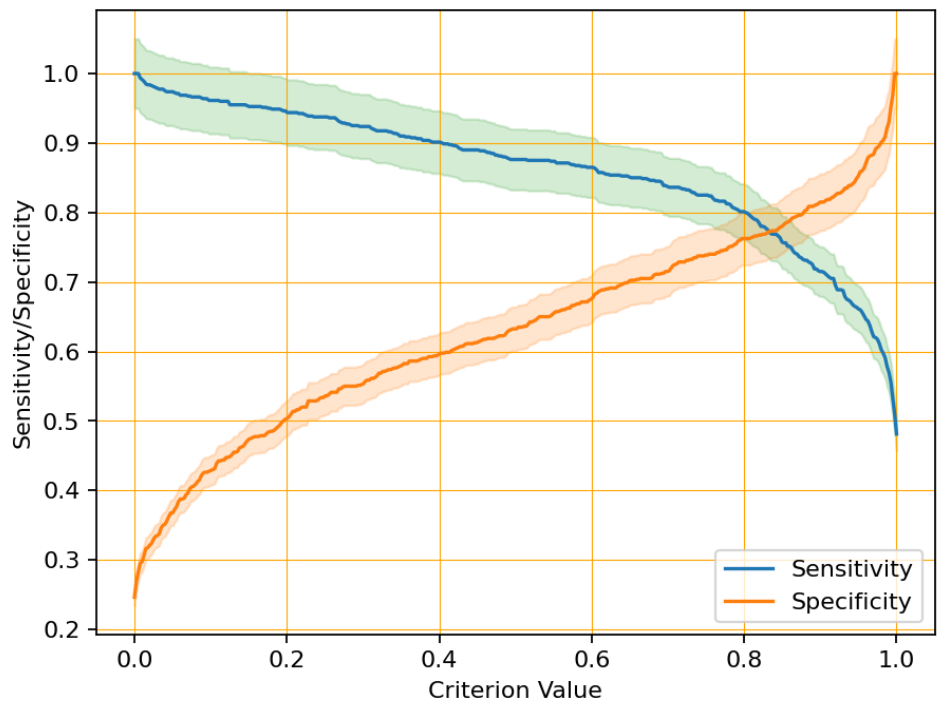

Supporting Figure 7: Sensitivity/Specificity curves of DenseNet backbone model. The orange curve denotes Specificity whereas blue curve denotes Sensitivity.

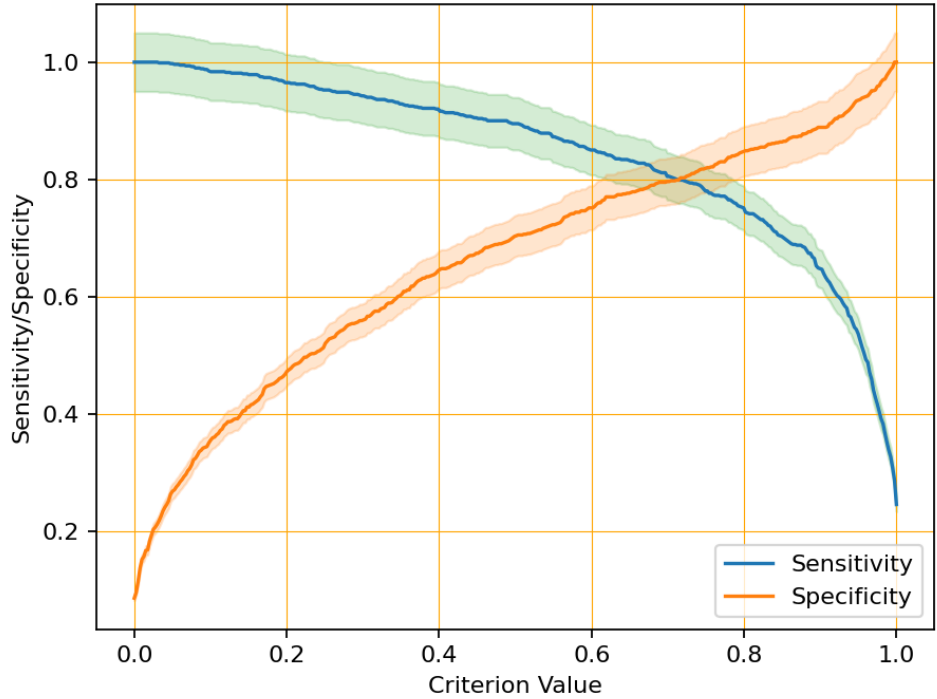

Supporting Figure 8: Sensitivity/Specificity curves of JointNet backbone model. The orange curve denotes Specificity whereas blue curve denotes Sensitivity. Since we apply our threshold to 0.6 in (-1,1) range, here it maps to 0.8 and which is proven with the graph above. The threshold ensures that JointNet do not overlook the active inflammation cases significantly.

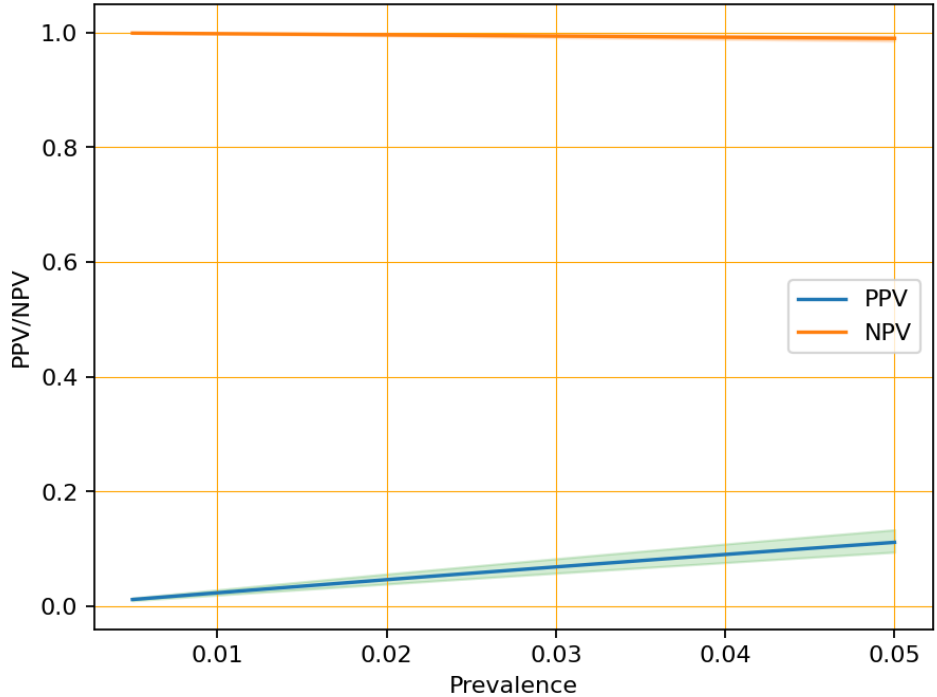

Supporting Figure 9: PPV and NPV of DenseNet and JointNet model against prevalence as prevalence between 0.0005 and 0.05. Prevalence thus impacts the positive predictive value (PPV) and negative predictive value (NPV) of tests. As the prevalence increases, the PPV also increases but the NPV decreases. Similarly, as the prevalence decreases the PPV decreases while the NPV increases. Since the model needs high PPV to not to overlook active inflammation patients.